\documentclass[%
 reprint,
superscriptaddress,
 amsmath,amssymb,
 aps,
prc,
]{revtex4-2}

\usepackage{graphicx}
\usepackage{dcolumn}
\usepackage{bm}
\usepackage{multirow}
\usepackage[dvipsnames]{xcolor}
\usepackage{rotating, graphicx}
\usepackage{float}
\usepackage{amssymb}
\usepackage{listings}
\usepackage{amsmath}
\usepackage{placeins}
\definecolor{airforceblue}{rgb}{0.36, 0.54, 0.66}
\definecolor{hotpink}{rgb}{1.00, 0.41, 0.70}
\definecolor{lawngreen}{rgb}{0.1, 0.5, 0}
\usepackage[colorlinks,bookmarks=false,
hypertexnames=true,allcolors=airforceblue]{hyperref}

\begin{document}

\preprint{APS/123-QED}

\title{Valence \boldmath$1s-0d$ proton vacancy of the \boldmath$^{32}$Si ground state}

\author{N.~Watwood}
 \email{nwatwood@anl.gov}
 \email{nwatwood@tamu.edu}
\author{C.~R. Hoffman}%
 \email{crhoffman@anl.gov}
\author{B.~P. Kay}
\author{I.~A. Tolstukhin}
\affiliation{%
Physics Division, Argonne National Laboratory, Lemont, Illinois 60439, USA
}%

\author{J.~Chen}
\affiliation{%
Department of Physics, Southern University of Science and Technology, 1088 Xueyuan Blvd, Nanshan, Shenzhen, Guangdong Province, China, 518055
}%
\author{T.~L.~Tang}
\affiliation{%
Physics Division, Argonne National Laboratory, Lemont, Illinois 60439, USA
}%
\author{ D.~Bazin}
\affiliation{%
FRIB/NSCL Laboratory, Michigan State University, East Lansing, Michigan 48824, USA}

\author{Y.~Ayyad}
\affiliation{%
IGFAE, Universidade de Santiago de Compostela, E-15782, Santiago de Compostela, Spain}

\author{S.~Beceiro-Novo}
\affiliation{%
Universidade da Coru$\Tilde{n}$a, Campus Industrial, Dpartamento de Fisica y Ciencias de la Tierra, CITENI, Ferrol, 15471, Spain}

\author{S.~J.~Freeman}
\affiliation{%
Department of Physics, University of Manchester, M13 9PL Manchester, United Kingdom}
\affiliation{EP Department, CERN, Geneva CH-1211, Switzerland}

\author{L.~P.~Gaffney}
\affiliation{%
Department of Physics, University of Liverpool, Liverpool L69 3BX, United Kingdom}

\author{R.~Garg}	
\affiliation{%
FRIB/NSCL Laboratory, Michigan State University, East Lansing, Michigan 48824, USA}

\author{H.~Jayatissa}
\affiliation{%
Physics Division, Argonne National Laboratory, Lemont, Illinois 60439, USA
}%

\author{A.~N.~Kuchera}	
\affiliation{%
Department of Physics, Davidson College, Davidson, North Carolina 28035, USA}

\author{P.~T.~MacGregor} 
\affiliation{%
Department of Physics, University of Manchester, M13 9PL Manchester, United Kingdom}

\author{A.~J.~Mitchell}
\affiliation{%
Department of Nuclear Physics and Accelerator Applications, Research School of Physics, Australian National University, Canberra ACT 2601, Australia
}

\author{A.~Mu\~noz-Ramos}
\affiliation{%
IGFAE, Universidade de Santiago de Compostela, E-15782, Santiago de Compostela, Spain}

\author{C.~Müller-Gatermann}
\affiliation{%
Physics Division, Argonne National Laboratory, Lemont, Illinois 60439, USA
}%
\author{F.~Recchia}	
\affiliation{%
Dipartimento di Fisica e Astronomia, Universitá degli Studi di Padova, I-35131 Padova, Italy}
\affiliation{
Istituto Nazionale di Fisica Nucleare, Sezione di Padova, I-35131 Padova, Italy
}

\author{C.~Santamaria}
\affiliation{Department of Physics and Engineering Physics, Morgan State University, Baltimore, Maryland 21251, USA. }

\author{M.~Z.~Serikow}
\affiliation{FRIB/NSCL Laboratory, Michigan State University, East Lansing, Michigan 48824, USA}

\author{D.~K.~Sharp}
\affiliation{%
Department of Physics, University of Manchester, M13 9PL Manchester, United Kingdom}

\author{G.~L.~Wilson}
\affiliation{Department of Physics and Astronomy, Louisiana State University, Louisiana 70803, USA}

\author{A.~H.~Wuosmaa}
\affiliation{%
Department of Physics, University of Connecticut, Storrs, Connecticut 06269, USA}

\author{J.~C. ~Zamora}	
\affiliation{%
FRIB/NSCL Laboratory, Michigan State University, East Lansing, Michigan 48824, USA}

\date{\today}

\begin{abstract}
The $^{32}$Si($^3$He\textit{,d})$^{33}$P reaction was studied in inverse kinematics at 6.3~MeV/$u$. States in $^{33}$P corresponding to the proton $1s-0d$ single-particle orbitals were identified up to $\sim$4.5 MeV in excitation energy. The ($^{3}$He,$d$) spectroscopic factors were determined from Distorted Wave Born Approximation calculations. When combined with complementary neutron-adding data, the $1s-0d$ proton vacancies in the $^{32}$Si ground state were extracted. In conjunction with a re-analysis of data from previous single-particle measurements, the trends in proton and neutron vacancy were explored across the $^{28,30,32,34}$Si isotopes. Both proton and neutron vacancy data show gradual changes in their occupancies. The proton $1s_{1/2}$ orbitals in $^{32}$Si and $^{34}$Si are both consistent with being empty. The ground-state nucleon distributions are described by shell-model calculations constrained to the $1s-0d$ model space.

\end{abstract}

\maketitle


\section{\label{sec:intro}Introduction}
A robust picture of a ground-state wave function is accessible through empirically derived single-particle occupancies. Such information can provide insight into the extent and impact of correlations, both those accounted for and those absent in a limited model space.
In the present case, new information gained from the $^{32}$Si ground-state occupancies enables a deeper exploration into the proton $0d_{5/2}-1s_{1/2}$ ($Z=14$) orbital spacing and its persistence as a sub-shell closure. Similarly for neutrons, the distribution of the two neutron holes within the traditional $N=20$ shell gap ($0d_{3/2}-0f_{7/2}$) informs on the size and type of ground-state correlations. 

Over a broader scope, ground-state nucleon occupancies across the even-A $Z=14$ isotopes are under examination for abrupt structural changes or, conversely, for evidence of smooth changes in the relative single-nucleon orbital energies. This avenue of exploration provides further insight into the $N\approx20$ transition region leading to $^{34}$Si, where discussions are ongoing about the evolving energy patterns of the $1p_{3/2}-1p_{1/2}$ spin-orbit partners~\cite{ref:Burgunder2014,ref:Mutschler2017,ref:Kay2017,ref:Orlandi2018,ref:Sorlin2021,ref:Chen2024,ref:Kuchera2024,ref:Chen2025front}.

An established method to extract the ground-state proton vacancy uses single-proton-adding transfer reactions. 
In cases where the ground-state total isospin is greater than zero ($T>0$) and its projection is equal to its total ($T_z = T$), the addition of a proton populates both lower and upper isospin states, $T_<$ and $T_>$, respectively~\cite{ref:French1961,ref:Wilkinson1969b}. 
The $T_<$ states reside at lower excitation energy and are populated directly in single-proton transfer. 
The $T_>$ states may also be populated directly; however, typical beam energies used for these reactions are often too low to populate all of these high-lying states.
Therefore, an often-adopted method is to utilize the single-neutron-adding data taken on the same ground-state which populates the isobaric analogs of the $T_>$ states. 

In the present work, this method is applied to determine the nucleon vacancies of the $0d_{5/2}$, $1s_{1/2}$, and $0d_{3/2}$ orbitals for the $^{32}$Si ground state. 
By all accounts, the mid-shell proton occupancy and nearly closed neutron occupancy of the $^{32}$Si ground state are well contained within the $1s-0d$ shell space~\cite{ref:For82,ref:Fornal1997,ref:Williams2023,ref:Heery2024,ref:Williams2025,ref:Williams2025b}. 
The analysis of the ground-state to ground-state two-neutron transfer data from $^{30}$Si($t$,$p$)$^{32}$Si supports this for the neutron case in particular~\cite{ref:For82}. 


\section{\label{sec:exp}Experiment}

The $^{32}$Si($^3$He\textit{,d})$^{33}$P single-proton adding reaction was studied in inverse kinematics using the SOLARIS solenoidal spectrometer and the ReA6 re-accelerator facility at the National Superconducting Cyclotron Laboratory (NSCL). A 6.3~MeV/$u$ long-lived $^{32}$Si beam was produced with an intensity of $\sim$10$^5$ particles per second. The beam was $\sim$90\% pure, with the dominant contaminant being the $^{32}$S isobar. The target consisted of a few $\mu$g/cm$^2$ of $^{3}$He thickness contained within a 400~$\mu$g/cm$^2$ titanium lattice. The $^{3}$He is present from the decay of the tritium (T$_{1/2}\approx12.3$~y) that had originally been loaded into the titanium $>15$ years prior. The experimental setup was contained within the uniform region of the 3~T magnetic field of the SOLARIS solenoidal spectrometer and paralleled the one described in Ref.~\cite{ref:Hoffman2012}.

Following a $(^3$He,$d$) reaction, the deuterons position and energy were measured upon their return to the beam axis by a four-sided array of position-sensitive silicon detectors (PSDs)~\cite{ref:Wuo07,ref:Lig10}. The array was positioned upstream of the target covering a relative distance between -570 cm~$\leq z \leq$ -220~cm. Projectile-like recoils were detected $80.5$~cm downstream of the target position by a silicon recoil telescope. The $\Delta$E-$E_{res}$ telescope consisted of two square (5x5 cm$^2$) silicon detectors with thicknesses of 53 $\mu$m and 150 $\mu$m, to measure the energy loss ($\Delta$E) and residual energy (E$_{res}$), respectively. The silicon telescope was shielded from high-rate, unreacted beam by a 12-mm diameter aluminum blocker centered on the beam axis. Timing signals from both the upstream PSD's and the downstream recoil silicon telescope provided a relative timing difference with a resolution on the order of FWHM~$\sim$20~ns.

\begin{figure}
    \centering
    \includegraphics[scale=0.55]{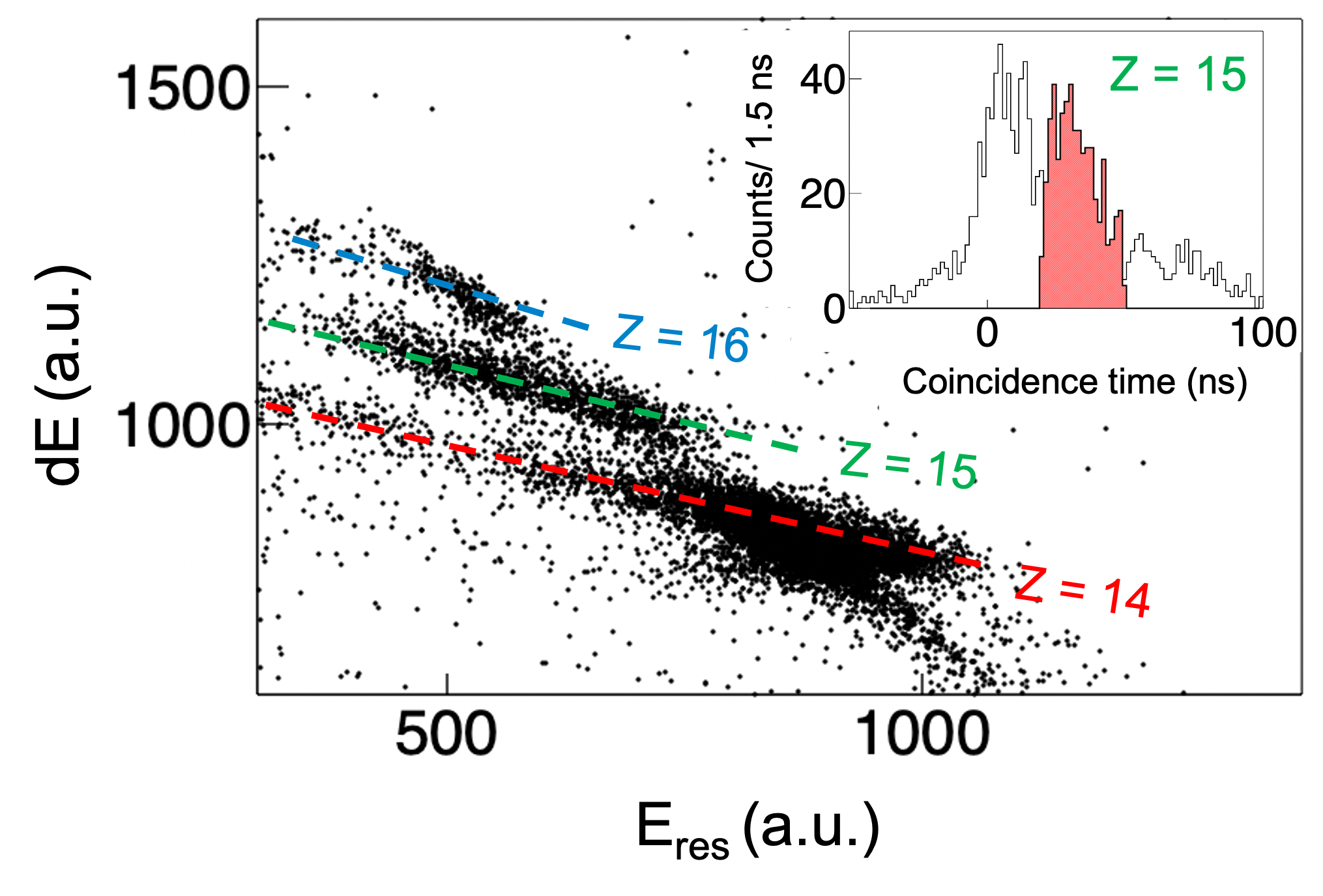}
    \caption{The heavy-ion recoil particle identification (PID) plot from the $\Delta$E-E$_{res}$ energies of the recoil silicon telescope. A condition requiring the detection of a charged particle by a PSD within 200~ns has been applied. The element groups having $Z=14$ (red), 15 (green), and 16 (blue) are labeled and indicated by overlaid dashed lines. The inset histogram shows the relative (coincidence) time between a heavy-ion recoil and a corresponding light charged particle for only $Z=15$ recoils. Calibration of the relative time difference was done such that protons are centered around 10~ns and deuterons are centered around $\sim30$~ns, shaded in red. 
    }
    \label{fig:dE_E}
\end{figure}

Selection of the $^{32}$Si($^3$He,$d$)$^{33}$P events relied on passing multiple data-selection conditions. Phosphorus recoils ($Z=15$) were identified by their energies in a particle identification plot (PID) of $\Delta$E vs. E$_{res}$ from the recoil telescope (Fig.~\ref{fig:dE_E}). A relative timing coincidence between the PSDs and the $\Delta$E-$E_{res}$ telescope provided discrimination of the light charged-particles through their cyclotron period times $T_{cyc}\propto A/q$~\cite{ref:Wuo07,ref:Lig10}. Selection of the single-orbit deuterons of interest was done by applying a 20-ns wide timing gate as shown in the inset of Fig.~\ref{fig:dE_E}. This timing gate differentiated protons produced by the simultaneously occurring $^{32}$Si($^3$He,$p$)$^{34}$P reaction, as well as other species of charged particles.

\section{Results}
\begin{figure}[t]
    \centering
    \includegraphics[scale=0.6]{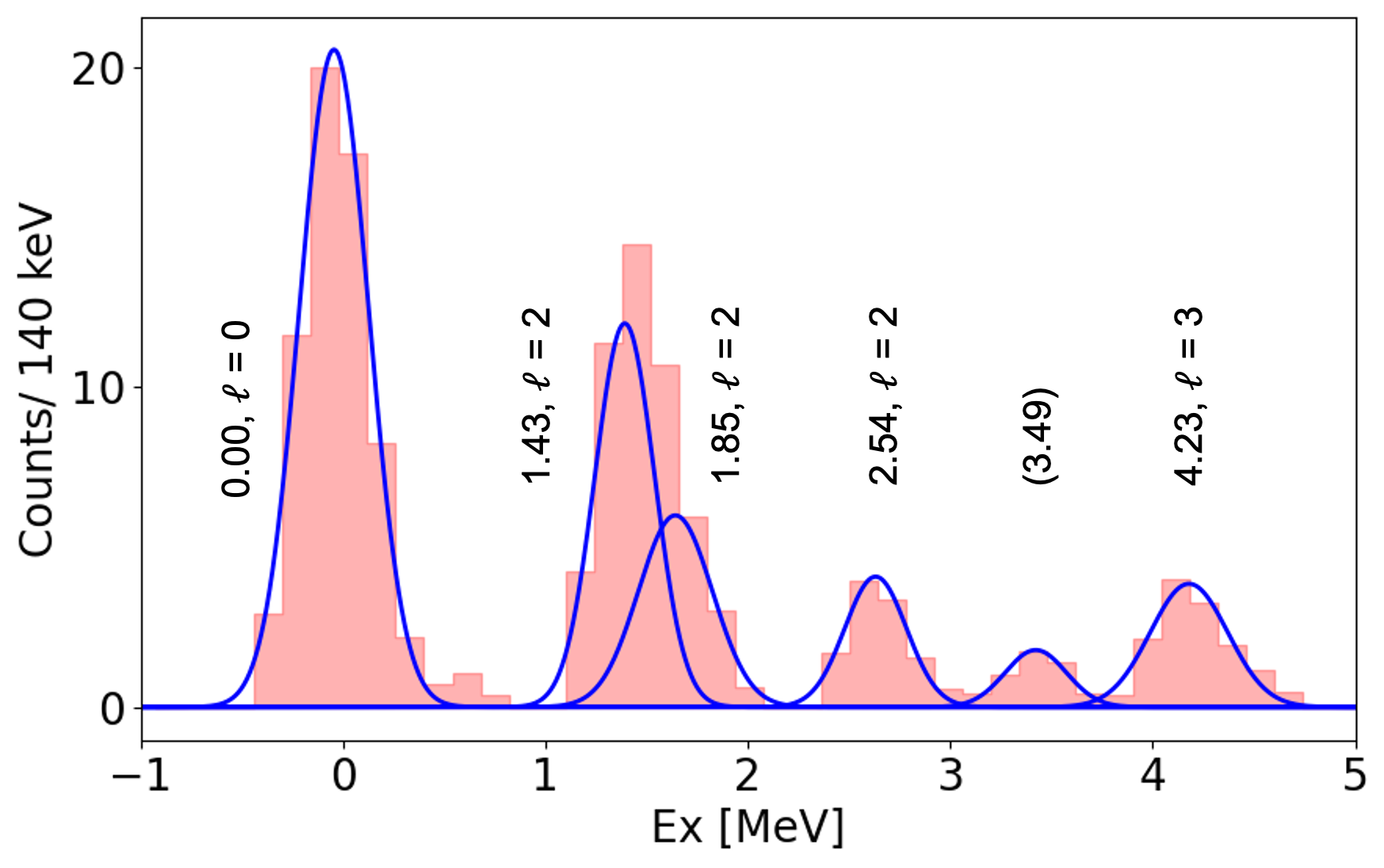}
    \caption{The excitation energy spectrum for states in $^{33}$P following the $^{32}$Si($^3$He,$d$) reaction at 6.3~MeV/$u$. States assigned to previously identified levels found in the literature are labeled by excitation and proton $\ell$ transfer. The unassigned level is labeled with its energy centroid in parenthesis.}
    \label{fig:32si_Ex}
\end{figure}

An excitation-energy spectra, $E_x$, for $^{33}$P was constructed from the energy and position information of deuterons passing both the PID and coincidence timing conditions (Fig.~\ref{fig:32si_Ex}). An $E_x$ ($Q$-value) resolution of approximately 470 keV FWHM was obtained, set primarily by the energy loss and straggling in the Ti lattice of the target. Six peaks were identified within the accepted $E_x$ range, $E_x$~$\lesssim4.5$~MeV. Four of these have been associated with known $1s-0d$ states at energies of $E_x = 0.000$~MeV (1/2$^+$), 1.431~MeV (3/2$^+$), 1.847~MeV (5/2$^+$), and 2.538~MeV(3/2$^+$)~\cite{33P_structure,ref:Lubna2018}. The orbital angular momentum of the transferred proton ($\ell$) and the associated orbital ($n\ell_{j}$) are fixed by the known spin-parity $J^{\pi}$ level assignments.

The state observed at 4.23 MeV agrees in energy with a known 7/2$^-_1$ level at $E_x=4.226$~MeV~\cite{ref:Grocutt2019}. This is likely the lowest-lying fragment from the 0$f_{7/2}$ proton orbit. A look at the $^{28,30}$Si$(^{3}$He,$d$) data, for instance, shows sizable single-proton overlaps to such a state~\cite{28Si_protonAdding,30Si_protonAdding}. The yield identified at 3.4(1)~MeV is in proximity to known levels at 3.276~MeV (3/2$^+$) and 3.491~MeV (5/2$^+$); however, an angular distribution was not possible due to low statistics. Therefore, the level was not assigned a $J^{\pi}$ and was excluded from the proton-vacancy determinations discussed below. An upper limit on a possible $\ell=2$ contribution to the $0d_{5/2}$ or $0d_{3/2}$ orbital vacancies based on this state was included in the uncertainties in their extracted vacancies. 

\begin{figure}
    \centering
    \includegraphics[scale=0.65]{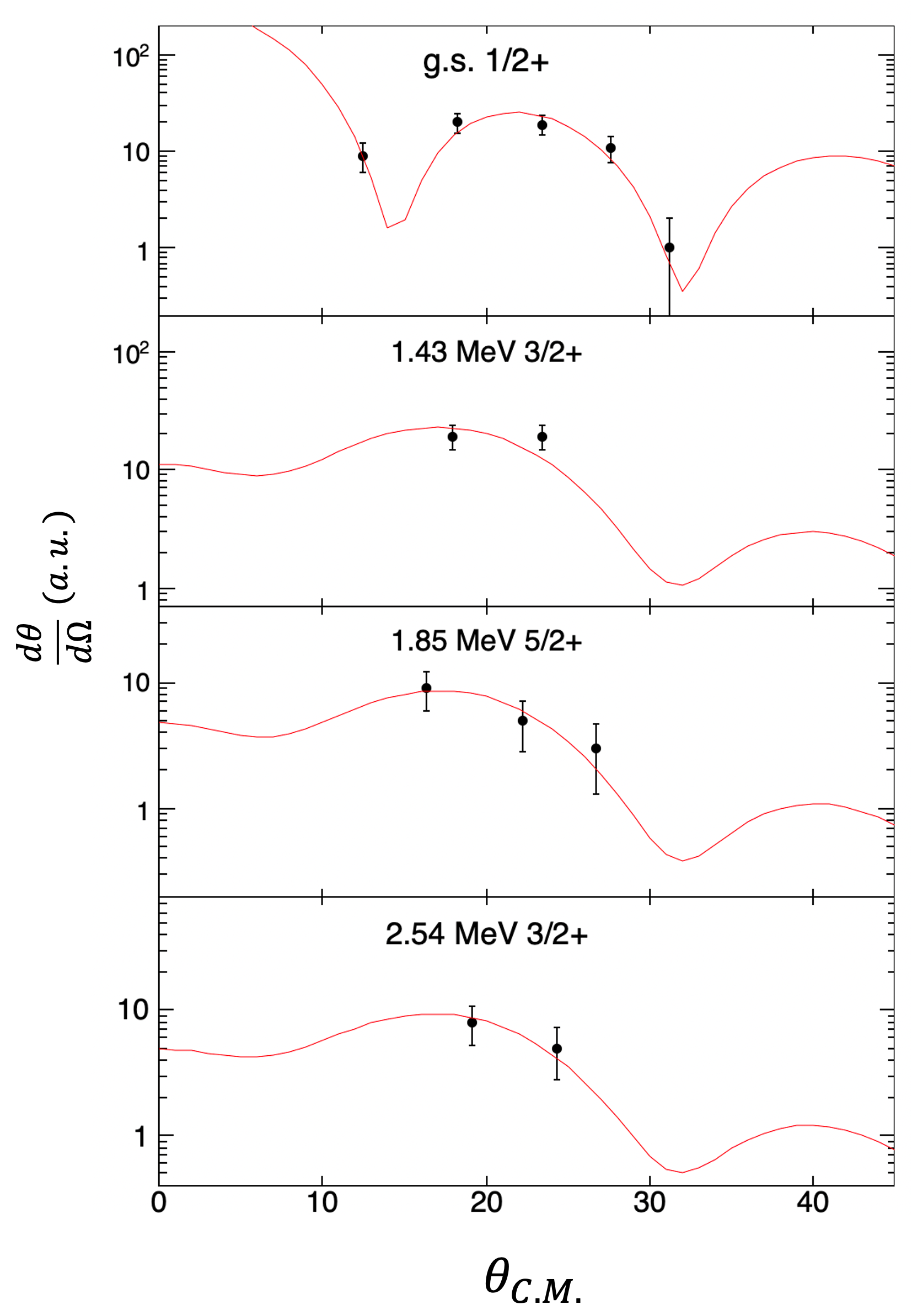}
    \caption{The measured ($^3$He,$d$) angular distributions for some of the observed states in $^{33}$P. The DWBA calculations for the expected $\ell$ transfer are also shown fit to the data. See text for details on the DWBA.}
    \label{fig:33P_angdist}
\end{figure}

The angular distributions for states where at least two angles were obtained are presented in Fig.~\ref{fig:33P_angdist}. 
Each angular bin corresponded to an equal length along the silicon array. The center-of-mass angle, $\theta_{\text{CM}}$, was calculated from the reaction kinematics for each excited state and the known physical location of the PSD's relative to the target position. The uncertainty of $\theta_{\text{CM}}$ is less than $0.5^{\circ}$. Since only relative strengths were required in the later analysis, no attempt was made to place an absolute scale for the differential cross sections, $d\sigma/d\Omega$. The dominant source of uncertainty on the relative $d\sigma/d\Omega$ was statistics, reflected in the data point error bars (Fig.~\ref{fig:33P_angdist}).

Distorted Wave Born Approximation (DWBA) calculations were used to extract spectroscopic factors, $C^2S$, for the overlaps between the $^{32}$Si ground-state and the observed final states in $^{33}$P. DWBA calculations were performed with the code PTOLEMY~\cite{ref:Mac78}. The optical-model parameters (OMPs) used for the incoming $^{3}$He and those for the outgoing deuteron were taken from the global models developed in Refs.~\cite{3He,deuteron}. The $^{3}$He-$d$ overlap form factor was calculated using the GFMC results of Ref.~\cite{ref:Brida2011}. The proton bound-state wave function in $^{33}$P was modeled by a Woods-Saxon potential in which the depth was varied in order to match the appropriate binding energy.  A bound-state radius $r_{0}=1.25$~fm and diffuseness parameter $a=0.65$~fm were used. Systematic uncertainties of 5\% were estimated in the relative $C^2S$ values determined for $\ell>0$ and 15\% for the $\ell=0$ transfer to the ground-state due to its increased sensitivity to both fitting and the optical-model-parameter values \cite{sum_rules}.

For $^{32}$Si, proton adding occurs on an isospin $T(T_z) = 2$ neutron-rich system, splitting the proton strength between the $T_<$ and $T_>$ isospin states. The strengths observed in $^{33}$P below $E_x$~$\approx4.5$~MeV are associated with the $T_<$ partners and $C^2S_{<} = 2T/(2T+1)S_{<} = \frac{4}{5}S_{<}$. The information on the upper $T_>$ isospin strengths taken from the neutron-adding data is presented below. We note that $C^2$ appears in the relation between the experimental cross sections to those calculated via the expression: 
\begin{equation}
    \frac{d\sigma(\theta)}{d\Omega} =  C^2S\frac{d\sigma(\theta)}{d\Omega}_{\text{DWBA}}
    \label{eq:dwba}
\end{equation}

The DWBA calculations are in good agreement with the $(^{3}$He,$d$) angular-distribution data, confirming the expected $\ell$ values of states based on their known $J^{\pi}$ (Fig.~\ref{fig:33P_angdist}). Although the angular acceptance was limited for the excited states, the key regions around the peaks in the $\ell=2$ distributions and much of the $\ell=0$ distribution were covered so as to provide reliable $C^2S$ extraction. The relative $C^2S_<$ values were deduced from the scaled values of the DWBA to the angular yields via Eq.~\ref{eq:dwba} and the corresponding relative $S_<$ values, substituting $C^2=4/5$, are listed in Table \ref{table:32Si_values}.

\begin{table*}[t]
\centering
\caption{Spectroscopic factors extracted from proton-adding ($^{3}$He,$d$) and neutron-adding ($d$,$p$) transfer reactions. The initial ground state is listed for each reaction. Each $S$ value was extracted from this work or the available data~\cite{28Si_protonAdding,ref:Peterson1983,30Si_protonAdding,30Si_neutronAdding,ref:Chen2024} using the DWBA approach and with the global optical-model parameter sets~\cite{3He,deuteron,KONING2003231}. The explicit correction for $C^2$ was included. The $S$ values have been normalized using the Macfarlane and French sum rules~\cite{Ptolemy} as described in the text. Spectroscopic factors are labeled by their final-state excitation energies ($E_x$) and spin-parities ($J^{\pi}$) as well as the corresponding orbital of the transferred light-particle ($n\ell_{j}$).}
\label{table:32Si_values}
\renewcommand{\arraystretch}{1.5}  
\begin{tabular}{cc@{\hspace{0.8cm}}c@{\hspace{0.8cm}}c@{\hspace{0.8cm}}c@{\hspace{0.8cm}}c@{\hspace{0.8cm}}c@{\hspace{0.8cm}}c@{\hspace{0.8cm}}c}
\hline
 & \multicolumn{4}{c}{($^3$He,\textit{d})} & \multicolumn{4}{c}{($d$,$p$)}\\
\hline
 & $E_x$ [MeV] & $J^{\pi}$ & $n\ell_{j}$ & $S$ & $E_x$ [MeV] & $J^{\pi}$ & $n\ell_{j}$ & $S$ \\
\hline
 \multirow{6}{*}{$^{28}$Si~\cite{28Si_protonAdding,ref:Peterson1983}} 
 & 0.00   & \( \frac{1}{2}^+ \) & $1s_{1/2}$ & 0.59(9)   &   0.00 & \( \frac{1}{2}^+ \) & $1s_{1/2}$ & 0.65(10) \\
 & 1.38    & \( \frac{3}{2}^+ \) & $0d_{3/2}$ & 0.82(4)   & 1.27 &  \( \frac{3}{2}^+ \) & $0d_{3/2}$ & 0.81(5)  \\
 & 1.95    & \( \frac{5}{2}^+ \) & $0d_{5/2}$ & 0.13(1)   &  2.03 & \( \frac{5}{2}^+ \) & $0d_{5/2}$ & 0.15(2)  \\
 & 2.42    & \( \frac{3}{2}^+ \) & $0d_{3/2}$ & 0.07(2)   &  2.43 & \( \frac{3}{2}^+ \) & $0d_{3/2}$ & 0.02(2)  \\
 & 3.10   & \( \frac{5}{2}^+ \) & $0d_{5/2}$ & 0.07(2)   &   3.07 & \( \frac{5}{2}^+ \) & $0d_{5/2}$ & 0.06(2) \\
 & 4.75    & \( \frac{1}{2}^+ \) & $1s_{1/2}$ & 0.04(2)   &  4.84 & \( \frac{1}{2}^+ \) & $1s_{1/2}$ & 0.09(5)  \\

\hline
\multirow{8}{*}{$^{30}$Si~\cite{30Si_protonAdding,30Si_neutronAdding}} 
& 0.00   & \( \frac{1}{2}^+ \) & $1s_{1/2}$ & 1.02(8)    & 0.00   & \( \frac{3}{2}^+ \) & $0d_{3/2}$ & 0.65(4) \\
& 1.27   & \( \frac{3}{2}^+ \) & $0d_{3/2}$ & 0.95(3)     & 0.75   & \( \frac{1}{2}^+ \) &$1s_{1/2}$ & 0.24(4) \\
& 2.23   & \( \frac{5}{2}^+ \) & $0d_{5/2}$ & 0.11(2)    & 2.32   & \( \frac{3}{2}^+ \) &$0d_{3/2}$ & 0.04(2)  \\
& 3.13   & \( \frac{1}{2}^+ \) & $1s_{1/2}$ & 0.04(2)      & 2.79   & \( \frac{5}{2}^+ \) & $0d_{5/2}$ & 0.06(2)  \\
& 3.51   & \( \frac{3}{2}^+ \) & $0d_{3/2}$ & 0.01(1)      & 4.26   & \( \frac{3}{2}^+ \) & $0d_{3/2}$ & 0.04(2) \\
& 4.19    & \( \frac{5}{2}^+ \) & $0d_{5/2}$ & 0.04(2)    & 4.72   & \( \frac{1}{2}^+ \) & $1s_{1/2}$ & 0.12(2)  \\
& 4.26   & \( \frac{3}{2}^+ \) & $0d_{3/2}$ & 0.01(1)    &    &  &      \\
& 4.59   & \( \frac{3}{2}^+ \) & $0d_{3/2}$ & 0.04(2)     &         &                &            \\

\hline\multirow{6}{*}{$^{32}$Si~\cite[this work,][]{ref:Chen2024}} & 0.00   
& \( \frac{1}{2}^+ \) & 1$s_{1/2}$ & 1.06(21) & 0.00      & \( \frac{3}{2}^+ \) & $0d_{3/2}$ & 0.37(4) \\
 & 1.43    & \( \frac{3}{2}^+ \) & 0$d_{3/2}$ & 0.67(15)  & 1.01 & \( \frac{1}{2}^+ \) & $1s_{1/2}$ & 0.25(5) \\
 & 1.85    & \( \frac{5}{2}^+ \) & 0$d_{5/2}$ & 0.16(5)   \\
 & 2.54    & \( \frac{3}{2}^+ \) & 0$d_{3/2}$ & 0.31(11)   \\
 & 3.4(1) & -- & -- & -- & \\
 & 4.23    & \( \frac{7}{2}^- \) & 0$f_{7/2}$ & $< 0.39$\footnote{Based on the yield at a single angle.}   \\
\hline\hline

\end{tabular}
\\ 
\end{table*}

The normalization procedure applied to the spectroscopic factors uses the Macfarlane and French sum rules~\cite{Ptolemy}. The proton vacancy across the entire proton $1s-0d$ shell was fixed equal to 6 based on the assumption of proton shell closures at $Z=8$ and $20$,
\begin{equation}
\begin{split}
    \sum_{n\ell_j} \langle \text{proton vacancy} \rangle_{n\ell_j} = 6, 
    \end{split}
    \label{eq:vacancies}
\end{equation}
where the $n\ell_j$ correspond to the $0d_{5/2}$, $1s_{1/2}$ and $0d_{3/2}$ orbitals. The proton vacancy for a single orbital ($n\ell_j$), accounting for distribution of the strength across both the upper and lower isospin projections, is given by
\begin{equation}
\begin{split}
    \langle \text{proton vacancy} \rangle_{n\ell_j} = G_p(T_<)_{n\ell_j} + G_p(T_>)_{n\ell_j}.\\ 
    \end{split}
    \label{eq:orbitalvacany}
\end{equation}
The total overlap strengths for each orbital, $G_p(T)_{n\ell_j}$, are then 
    \begin{equation}
\begin{split}
     G_p(T_<)_{n\ell_j} = \sum (2j+1)C^2S_{<} = \sum (2j+1)\frac{2T}{2T+1}S_{<}, \\
    \end{split}
    \label{eq:protonstrength}
\end{equation}
and 
    \begin{equation}
\begin{split}
     G_p(T_>)_{n\ell_j} = \sum (2j+1)\frac{1}{2T+1}S_{>} \\
     = \frac{1}{2T+1}\langle \text{neutron vacancy} \rangle_{n\ell_j}.\\
    \end{split}
    \label{eq:neutronstrength}
\end{equation}
In the expressions above, $J=0$ is used as the initial state spin and the $C^2$ values have been explicitly written in terms of the initial total isospin, $T$. The summations imply the inclusion of all fragments of strength across the various states belonging to that particular orbital.

The $T_<$ states of interest reside within the $E_x$ energy range probed directly in the proton-adding reaction. The $T_>$ states occur typically at $E_x$~$\gtrsim S_p$ in these nuclei and are more challenging to directly populate in proton-adding reactions. The $S_{>}$ values are taken instead from single-neutron-adding spectroscopic factors as they are the isobaric analogues~\cite{ref:French1961,ref:Wilkinson1969b}. The summation in Eq.~\ref{eq:neutronstrength} is then equivalent to the neutron vacancy in the $T_>$ system, scaled by the appropriate $C^2$ factor. The neutron-adding 3/2$^+$ ground state [$S=0.37(4)$] and the 1.01-MeV 1/2$^+$ state [$S=0.25(5)$] $S_>$ information was gathered from the $^{32}$Si(\textit{d,p}) reaction data of Ref.~\cite{ref:Chen2024} (Table~\ref{table:32Si_values}). The neutron-adding $S_>$ were independently normalized using the Macfarlane and French sum-rule procedures to a total summed strength of 2, the number of $1s-0d$ neutron holes within the $N=20$ closed shell.

Inserting the appropriate proton and neutron $1s-0d$ shell vacancy values, as well as the proper isospin values into Eqs. (\ref{eq:orbitalvacany}) - (\ref{eq:neutronstrength}), gave the expression for the normalization of the $S_<$. Normalized $S$ values are shown in Table~\ref{table:32Si_values}. Table~\ref{table:spectroscopic_values} gives the corresponding $G_p$ values separated by the upper and lower components. The derived proton vacancy for each $1s-0d$ proton orbital in $^{32}$Si is in the final column of Table~\ref{table:spectroscopic_values}. The proton vacancy uncertainties are dominated by the errors on the $S_<$ values and also include a systematic uncertainty due to possible missing strengths outside of the accepted $E_x$ regions ($\lesssim5$\%).

\begin{table}[t]
    \centering
     \caption{The extracted $1s-0d$ proton vacancy values and a breakdown of their isospin lower and upper contributions in $^{28,30,32}$Si. The lower and upper spectroscopic strengths are taken from Eqs.~\ref{eq:protonstrength} and~\ref{eq:neutronstrength}. The proton vacancies for $^{34}$Si are determined from the extracted proton occupancies of Ref.~\cite{ref:Mutschler2017}. 
    \label{table:spectroscopic_values}}
    \renewcommand{\arraystretch}{1.5}  
    \begin{tabular}{cccccc}
    \hline
     & T & $n\ell_j$ & $G_p(T_<)$ & $G_p(T_>)$ & $\langle\text{proton vacancy}\rangle_{n\ell_j}$ \\
     \hline
     \multirow{3}{*}{$^{28}$Si} & \multirow{3}{*}{0}  
       & $0d_{5/2}$ & 1.2(1) & --  & 1.2(1)\\
     & & $1s_{1/2}$ & 1.3(2) & --  & 1.3(2)\\
     & & $0d_{3/2}$ & 3.6(3) & --  & 3.6(3)\\
    \hline
    \multirow{3}{*}{$^{30}$Si} & \multirow{3}{*}{1}  
       & $0d_{5/2}$ &  0.6(1) & 0.1(1)  & 0.7(1)\\
     & & $1s_{1/2}$ &  1.4(2) & 0.2(1)  & 1.7(2)\\
     & & $0d_{3/2}$ &  2.7(2) & 1.0(1)  & 3.7(2)\\
    \hline
     \multirow{3}{*}{$^{32}$Si} & \multirow{3}{*}{2}  
       & $0d_{5/2}$ & 0.8(3) & 0.0(1)  & 0.8(3)\\
     & & $1s_{1/2}$ & 1.7(3) & 0.1(1)  & 1.8(3)\\
     & & $0d_{3/2}$ & 3.3(7) & 0.3(1)  & 3.6(7)\\
     \hline
    \multirow{3}{*}{$^{34}$Si} & \multirow{3}{*}{2}  
       & $0d_{5/2}$ & -- & --  & 0.5(5) \\
     & & $1s_{1/2}$ & -- & --  & 1.8(2) \\
     & & $0d_{3/2}$ & -- & --  & 3.7(2) \\
     \hline
    \end{tabular}
\end{table}

\section{Discussion}

\subsection{Correlations in the $^{32}$Si Ground State\label{sec:corr}}

The extracted proton vacancies over the $1s-0d$ orbitals for $^{32}$Si are shown in red in Fig.~\ref{fig:proton_occupancies}. 
From the independent-particle model, the expected vacancies for these orbitals are: $0d_{5/2}$=0, $1s_{1/2}$ = 2, and $0d_{3/2}$ = 4. These limits are shown in Fig.~\ref{fig:proton_occupancies} by the thicker black lines. The $1s_{1/2}$ and $0d_{3/2}$ orbitals are both consistent with being empty orbitals. Lower limits on their vacancies are $\gtrsim 1.5$ and $\gtrsim2.9$, respectively, based on their vacancy errors. A proton vacancy of $0.8(3)$ for the $0d_{5/2}$ orbital is consistent with no more than $\approx$~1 nucleon being excited out of this orbital and distributed across the upper two orbitals. 

The new proton-vacancy information posits $^{32}$Si as having a robust $Z=14$ proton sub-shell as vacancies are close to the independent-particle-model expectations with $\lesssim1$ proton occupying the combined $1s_{1/2}$ and $0d_{3/2}$ orbitals. This agrees with interpretations built out of some recent results~\cite{ref:Heery2024,ref:Williams2023,ref:Chen2024,ref:Williams2025,ref:Williams2025b}. For example, the presence of a $5^-$ (yrast) spin-trap isomer at $E_x = 5.505$~MeV is in part due to the larger proton excitation energy requirement needed to generate the yrast $4^+$ level relative to the other nearby $N=18$ isotones (see Fig. 6 of Ref.~\cite{ref:Williams2023}). The excitation energy of the $4^+_1$ level, as well as its B($E2$) strength to the $2^+_1$ level, were both reproduced by shell model calculations using the $1s-0d$ constrained USDB interaction~\cite{ref:Bro06}.

The $0^+_1\rightarrow2^+_1$ B($E2$) transition strength has been recently determined using \emph{safe} Coulomb excitation of $^{32}$Si on Pt [143(20)~$e^2$fm$^4$] ~\cite{ref:Heery2024} and from a lifetime measurement of the $2^+_1$ level [189(29)~$e^2$fm$^4$]~\cite{ref:Williams2025b}. These are in addition to reliable past results of 160(60)~$e^2$fm$^4$~\cite{ref:Pronko1972} and 113(33)~$e^2$fm$^4$~\cite{ref:Ibb98}. The recent Coulomb excitation value, which overlaps with both previous results, was found to be 15\% below the predictions of the USDB interaction calculations~\cite{ref:Heery2024,ref:Bro06}. This level of disagreement would be a unique situation in the region, as shown systematically in Fig.~4 of Ref.~\cite{ref:Heery2024}, and would suggest reduced contributions from correlations in $^{32}$Si, such as core-polarization in or between the two active states. However, the value from the $2^+_1$ lifetime suggests an increase in deformation compared to the Coulomb excitation results and is more in-line with the theoretical calculations (Table VI of Ref.~\cite{ref:Williams2025b}).

Finally, additional evidence for reduced correlations in the $^{32}$Si ground state was found from the weakly fragmented $^{33}$Si single-neutron strength distribution determined in Ref.~\cite{ref:Chen2024}. A majority of the single-neutron strength for the $0f_{7/2}$ and $1p_{3/2(1/2)}$ orbitals was contained within a single energy state. This is in contrast to neighboring $N=18$ isotones, $^{34}$S~\cite{ref:Kuchera2024} and $^{36}$Ar~\cite{ref:Sen1974}, where the neutron $1p$ strength was more evenly distributed across two states. Shell-model calculations allowing for neutron excitations into the $0f-1p$ shell but still utilize the USDB interaction for the proton orbitals reproduce the lack of fragmentation~\cite{ref:Lubna2020}. 

\subsection{Proton \& Neutron Vacancy Along \boldmath$Z=14$}

The $1s-0d$ proton vacancies for the $^{28,30,34}$Si isotopes were extracted to explore their evolution along the $Z=14$ isotopic chain. Vacancies for $^{28,30}$Si were determined from the data in the literature using the same procedures detailed above for $^{32}$Si. The single-proton adding data on $^{28,30}$Si were taken from ($^{3}$He\textit{,d}) measurements at 6.15~MeV/$u$~\cite{28Si_protonAdding} and 8.33~MeV/$u$~\cite{30Si_protonAdding}, respectively. For $^{28}$Si ($T=0$), there is only one accessible isospin projection. The single-proton and single-neutron states are analogs of each other. Only the proton-adding spectroscopic factors were used and normalized to the total $1s-0d$ proton vacancy using $C^2=1$. In the case of $^{30}$Si, the $S_>$ strengths were taken from one neutron-adding ($d$,$p$) reaction data~\cite{30Si_neutronAdding} and were independently normalized to the $1s-0d$ shell neutron vacancy of 4.

\begin{figure}[t]
     \centering
     \includegraphics[width=0.48\textwidth]{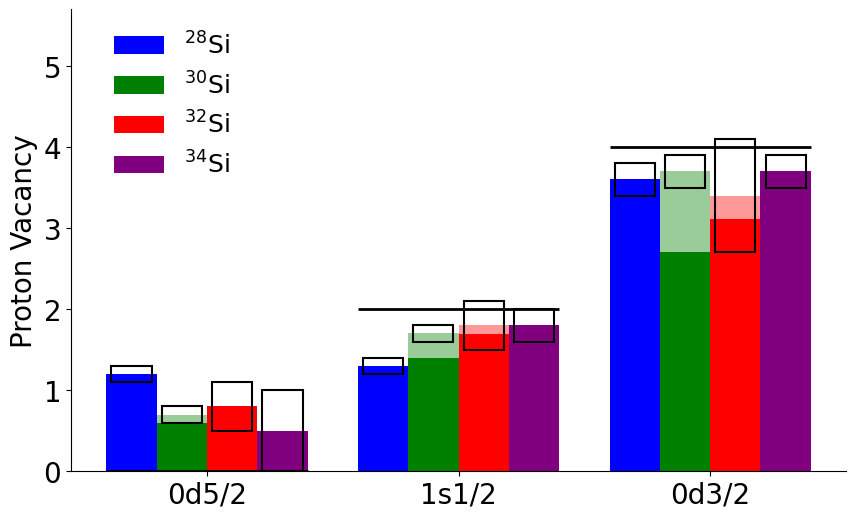}
     \caption{Extracted proton vacancies for the even-$A$ Si isotopes from $A=28-34$ over the $1s-0d$ orbitals (Table~\ref{table:spectroscopic_values}). The lighter shaded regions indicate the contributions from the $T_>$ states to the vacancy. The black rectangles indicate the error on the extracted proton vacancy. The thick black lines indicate the independent single-particle-model proton-vacancy expectations.}
     \label{fig:proton_occupancies}
\end{figure}


The previously measured proton- and neutron-adding cross sections were reanalyzed through the DWBA approach using the same bound-state formulations and global optical models as used in the $^{32}$Si analysis~\cite{3He,deuteron,KONING2003231}. The normalized $S$ values determined from the reanalysis of $^{28}$Si and $^{30}$Si are shown in Table~\ref{table:32Si_values} and they are in agreement with the previously extracted values within uncertainties. Only states contributing $G_p \gtrsim 0.05$ were included in the determination of proton vacancies. An arbitrary cutoff at $E_x$~=~5~MeV was used for the extracted $T_<$ strengths. This ensured no contributions from direct population of $T_>$ analog states. It also included $>95\%$ of the available strength to the states of interest based on the data re-analyzed by the $^{28,30}$Si single-particle adding data. The uncertainties of $S$ include the DWBA fit uncertainties and the same systematic uncertainties prescribed for states in the $^{32}$Si($^{3}\text{He},d)$ analysis. 

The proton vacancies for the $0d_{5/2}$, $1s_{1/2}$ and $0d_{3/2}$ orbitals are listed in Table~\ref{table:spectroscopic_values} and Figure \ref{fig:proton_occupancies} for the $Z=14$ even-even isotopes from $A=28 - 34$. The additional data for $^{34}$Si were taken from Ref.~\cite{ref:Mutschler2017}. In this work, single-proton knockout yields were analyzed to provide the relative proton occupancies in $^{34}$Si. The occupancies for the $1s-0d$ orbitals were normalized to the 6 protons within the shell. Vacancies were determined through subtraction of the extracted occupancies from the full $2j+1$ occupancies of each orbital. An additional uncertainty of $\sim0.1$ nucleons was added to account for up to $5$\% of total strength possibly residing above the $E_x=5$~MeV cutoff.

\begin{figure}[t]
     \centering
     \includegraphics[width=0.48\textwidth]{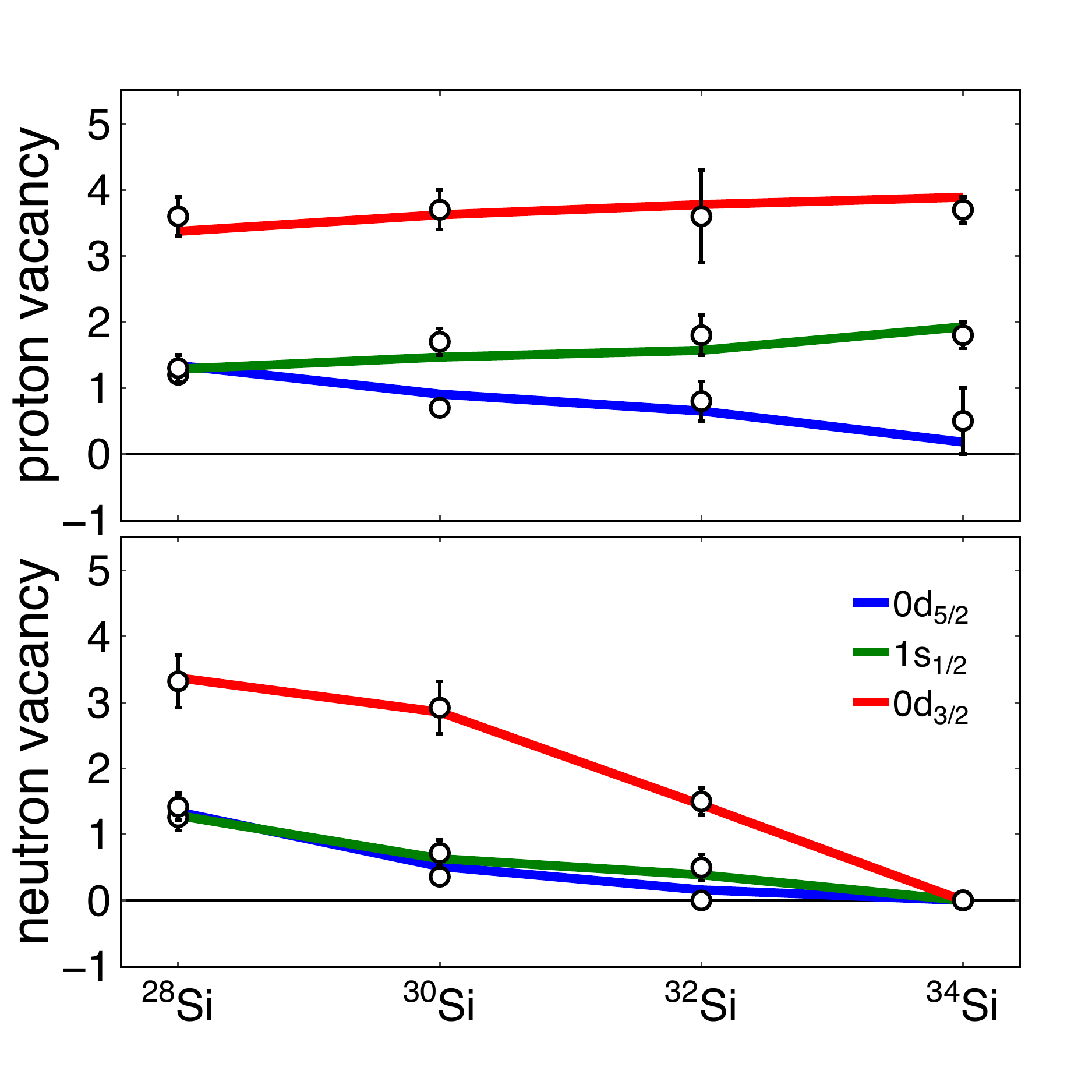}
     \caption{A comparison of the empirical proton and neutron vacancies (data points) to those based on configuration-interaction shell-model calculations using the USDB $1s-0d$ effective interaction (lines)~\cite{ref:Bro06}. The normalization procedure applied to the data is given in the text. 
     }
     \label{fig:theory}
\end{figure}

For all isotopes considered here, the independent single-particle limits have a fully occupied $0d_{5/2}$ orbital and vacant $1s_{1/2}$ and $0d_{3/2}$ orbitals. Within the $1s-0d$ space, any occupancy of the $1s_{1/2}$ or $0d_{3/2}$ orbitals must deplete the occupancy of the $0d_{5/2}$ orbital. This effect is qualitatively observed through the slight \emph{decrease} in the vacancy of the $0d_{5/2}$ orbital with increasing $A$, compensated mostly by an \emph{increase} in the vacancy of the $1s_{1/2}$ orbital.

The move towards a filled $0d_{5/2}$ proton orbital with increased filling of the neutron orbitals follows the reduction of the B($E2$) transition strengths, as shown in Fig.~4(a) of Ref.~\cite{ref:Heery2024}. Both observables are indications of a (smooth) reduction in the deformation or particle-hole contributions, approaching the $N=20$ shell gap in these ground states. Also, the 
proton $0d_{3/2}$ orbital remains $\gtrsim90$\% vacant and constant across all of the isotopes within errors of $\approx0.3$ nucleons.

To complement the proton data, the neutron $1s-0d$ vacancies were also obtained from the neutron-adding data available in Table~\ref{table:32Si_values} for $^{28}$Si~\cite{ref:Peterson1983}, $^{30}$Si~\cite{30Si_neutronAdding} and $^{32}$Si~\cite{ref:Chen2024}. The single-neutron adding data on $^{34}$Si showed no evidence for vacancy within $1s-0d$ shell~\cite{ref:Burgunder2014}. The neutron-vacancy information is shown alongside the proton values in Fig.~\ref{fig:theory}. The decrease in the neutron vacancy with $A$ occurs as expected due to the filling of the $0d_{5/2}$ at $N=14$, followed by the filled $1s_{1/2}$ at $N=16$, and the sequential filling of the $0d_{3/2}$ at $18$ and $20$. Similar to the proton occupancies, the distributions of neutrons continues to align better with the normal order filling as $N=20$ is approached. On average, $\approx1$ neutron is displaced outside the expected $1s_{1/2}$ closed shell in $^{30}$Si, reducing to $\lesssim0.5$ neutrons in $^{32}$Si, and to first order $\approx0$ in $^{34}$Si

\subsection{Relevance to the Evolution of the Neutron Spin-Orbit Partner Energy Differences}

There have been discussions surrounding the nature of the evolution of the energy difference between the \emph{neutron} $1p_{3/2}-1p_{1/2}$ spin-orbit partners in the $N=20$ region~\cite{ref:Duguet2017,ref:Mutschler2017,ref:Kay2017,ref:Orlandi2018,ref:Sorlin2021,ref:Chen2024}. Some authors deduce a sudden reduction in this energy difference between $^{36}$S and $^{34}$Si to be attributed to a substantial change in their interior proton density. Specifically, the removal of the protons from a nearly full $1s_{1/2}$ proton orbital in $^{36}$S to a nearly vacant $1s_{1/2}$ orbital in $^{34}$Si results in a weakening of the spin-orbit splitting.

A counterpoint to the above has been made based on the observation of continuous trends in the $1p_{3/2} - 1p_{1/2}$ single-neutron centroid energy separation throughout the $Z\approx12 - 20$, $N\approx16-20$ region~\cite{ref:Kay2017,ref:MacGregor2021,ref:Chen2024,ref:Chen2025front} which included energies from $^{31,33}$Si and $^{35}$Si. One key aspect in deducing these trends was the proper inclusion of all known single-neutron fragments in the energy centroids. Within this same picture, a natural explanation also unfolds for $^{29}$Mg which has the smallest measured energy separation of $\sim1.0$~MeV.


No definitive correlation between the $1p_{3/2}-1p_{1/2}$ single-neutron energy spacing and the proton $1s_{1/2}$ vacancy is found in the available $Z=14$ data. The relative energy spacing is $\approx1.5-1.6$~MeV 
for $^{31}$Si and $^{33}$Si and it reduces to $\approx1.1$~MeV 
in $^{35}$Si as shown in Fig.~4 of Ref.~\cite{ref:Chen2024}. Although a change in energy spacing occurs between $^{32}$Si and $^{34}$Si, both isotopes are, within their errors, consistent with a fully vacant $1s_{1/2}$ proton orbital. Similarly, $^{30}$Si too, is limited to a small $1s_{1/2}$ proton occupancy ($\lesssim0.5$ nucleons). The proton $0d$ orbital vacancies are also in agreement with each other across these three isotopes. Only $^{28}$Si stands out as having a larger $0d_{5/2}$ proton vacancy and a correspondingly increased $1s_{1/2}$ proton occupancy.


\subsection{Comparisons with $1s-0d$ Shell Model Calculations\label{sec:calcs}}

Shell-model calculations constrained to the $1s-0d$ shell were carried out using the USDB interaction~\cite{ref:Bro06}. We note that this interaction makes up the $1s-0d$ model space of the FSU interaction~\cite{ref:Lubna2020}, which has described much of the level energy, transition strength and single-neutron $0f_{7/2}$ and $1p$ data available on $^{32}$Si~\cite{ref:For82,ref:Fornal1997,ref:Williams2023,ref:Heery2024,ref:Williams2025,ref:Williams2025b,ref:Chen2024}. The calculated ground-state proton- and neutron-occupancy values were extracted and the corresponding orbital vacancies are shown in Fig.~\ref{fig:theory} by the solid lines. The vacancies are in agreement with the experimental data for both protons and neutrons across all of the isotopes. In particular, reproduction of the gradual trends in proton occupancy across the $0d_{5/2} - 1s_{1/2}$ orbitals is found. Similarly, the neutron trends are well described as the orbitals are filling. The calculations have a strong grasp of the ground-state wavefunctions in these $Z=14$ isotopes.

Despite reproducing the proton ground-state occupancies in $^{32}$Si, there are indications that the USDB calculations fail to reproduce the magnitude of the B($E2$) to the first excited $2^+$ state by over 15\% if the recent Coulomb excitation energy data is adopted~\cite{ref:Heery2024}. This is not the case, however, if the lifetime data of Ref.~\cite{ref:Williams2023,ref:Williams2025,ref:Williams2025b} is taken. There, though the measured B($E2$) value is also below the theoretical number, they agree within the upper limit on the experimental error. If the experiment-to-theory discrepancy persists through future scrutiny, one resolution may be the need for an inclusion of forces, residing outside the $1s-0d$ model space, that were unable to be captured in the USDB interaction formation. Such forces may originate coherently from within the lower $0p$ shell and manifest as changes in the core polarization and in the effective charge values. We note that typical values of $e_p = 1.35-1.35e$ and $e_n = 0.35-0.45e$ were used in the calculated results of Refs.~\cite{ref:Heery2024,ref:Williams2025b}. In addition, the same values used with the FSU (USDB) interaction did give a consistent description of the $4^+_1 \rightarrow 2^+_1$ B($E2$) transition strength~\cite{ref:Williams2023,ref:Williams2025,ref:Williams2025b}. 

An alternative scenario, if the discrepancy persists, is that the configurations of the $2^+_1$ level in $^{32}$Si are the main contributing factor. Namely, that perhaps some small but impactful configurations vary between the $0^+_1$ and $2^+_1$ wave functions. To date, no single-particle reaction data into or from this state is available. The $^{30}$Si($t$,$p$) cross sections to the $2^+_1$ level were poorly reproduced by the shell-model interaction used in Ref.~\cite{ref:For82}. As stated in that paper, reproducing the relative yields of $2^+$ states populated in ($t$,$p$) were a challenge in the region due to significant changes in cross sections resulting from small contributions of intruder ($0f-1p$) configurations.

\section{Summary}
The $^{32}$Si ground-state single-proton occupancies extracted from ($^3$He,$d$) yields showed a small vacancy in the $0d_{5/2}$ orbital ($\lesssim 1$), and similarly, nearly vacant $1s_{1/2}$ and $0d_{3/2}$ proton orbitals. The distribution of the protons across the $1s-0d$ orbitals in $^{32}$Si are the same as those for $^{34}$Si within uncertainties. Overall, small changes in both the proton and neutron occupancies were observed when moving from $^{30}$Si to $^{34}$Si, in particular. The evolution of the occupancies was in line with expectations based on the approach towards the $N=20$ shell closure. Both the ground-state proton and neutron occupancies for each of the even-$A$ Si isotopes investigated are reproduced by the calculated occupancies of the USDB interaction.

\section*{Acknowledgments}
This work was supported by the U.S. Department of Energy, Office of Science, Office of Nuclear Physics, under Contract Nos. DE-AC02-06CH11357 (Argonne) and DE-SC0014552, by the UK Science and Technology Facility Council via Grant Ref. ST/T004797/1, ST/V001116/1 and ST/Y000323/1, by the International Technology Center Pacific (ITC-PAC) under Contract No. FA520919PA138 and Australian Research Council Grant No. DP210101201, and by Grant RYC2020-030669 and PID2022-142557NA-I00 funded by MCIN/AEI/ 10.13039/501100011033, FSE+ and FEDER, UE. This material is based upon work supported by NSF’s National Superconducting Cyclotron Laboratory which was a major facility fully funded by the National Science Foundation under award PHY-1565546; SOLARIS is funded by the DOE Office of Science under the FRIB Cooperative Agreement DE-SC0000661. J. Chen was supported by the National Natural Science Foundation of China under Contracts No. 12475120 and No. 12435010. Y.A. is supported by grant RYC2019-028438-I and PID2021-125995NA-I00 funded by MCIN/AEI/10.13039/501100011033 and by the Regional Government of Galicia under the program “Proyectos de excelencia” Grant No. ED431F 2022/13.

\bibliography{apssamp,raisorbib,references}

\end{document}